\documentclass[12pt]{article}
\usepackage[utf8]{inputenc}

\title{Optimal Reduction without Oracle?}
\author{Anton Salikhmetov}

\usepackage[utf8]{inputenc}
\usepackage[english]{babel}
\usepackage[title]{appendix}
\usepackage{fullpage}
\usepackage{csquotes}
\usepackage[
    backend=bibtex,
    style=numeric,
    block=ragged,
    url=false,
    sorting=none,
    firstinits
]{biblatex}
\usepackage{amsmath}
\usepackage{amssymb}
\usepackage{amsthm}
\usepackage{fixmath}
\usepackage{wrapfig}
\usepackage{caption}
\usepackage{subcaption}
\usepackage{tikz}
\usepackage{tikz-inet}

\tikzset{every node/.style = {node distance=0em, scale=0.8}}

\newcommand\fan[2]{\delta^{\,#2}_{#1}}

\begin{document}
\maketitle

\begin{abstract}
We suggest an approach to solve the problem of matching fans in interaction net implementations of optimal reduction for the pure untyped lambda calculus without use of any additional agent types.
Our implementation supports a wider class of lambda terms than the abstract version of Lamping's algorithm and beats the interaction net implementation of closed reduction by the total number of interactions.
\end{abstract}

\section{Problem}

Matching fans is the main problem of implementation of optimal reduction in interaction nets~\cite{optimal}.
Existing solutions use so-called oracle which is implemented using bracket and croissant agents in BOHM~\cite{optimal} and delimiter agents in Lambdascope~\cite{lambdascope}.
Unfortunately, both versions produce significant overhead.

Since that overhead is due to the oracle, we decided to limit our signature to the basic types only (abstraction, application, erase, and fan) and experiment with more sophisticated structures attached to fans rather than just integer indices, aiming at the same behavior of agents as if the oracle were still present.

Our experiments resulted in a partial solution we present in this paper.
It benefits from optimal number of $\beta$-reductions like BOHM and Lambdascope.
Simultaneously, it has performed better by the number of interactions than non-optimal CVR~\cite{mackie}.

\section{Solution}

We work in interaction calculus~\cite{calculus}.
The signature of our interaction system is
$$
\Sigma = \{@, \lambda, \varepsilon\} \cup \{\fan t n\ |\ t \in \mathbb{T},\ n \in \mathbb{N}\},
$$
elements of $\mathbb{T}$ being inductively defined as two-colored binary trees with names in each leaf:
$$
t ::= x\ |\ \langle t_1 + t_2 \rangle\ |\ \langle t_1 - t_2 \rangle.
$$

Intuitively, one can think of $n$ in $\fan t n$ as number of this agent's instances created during its life, and $t$ can be thought of as this agent's identity.

The interaction rules are as follows:
\begin{align*}
\varepsilon &\bowtie \alpha[\varepsilon,\dots,\ \varepsilon] \quad (\forall \alpha \in \Sigma); \\
@[x,\ y] &\bowtie \lambda[x,\ y]; \\
\fan t n[\lambda(x,\ y),\ \lambda(v,\ w)] &\bowtie \lambda[\fan t {n + 1}(x,\ v),\ \fan t {n + 1}(y,\ w)]; \\
\fan t n[@(x,\ y),\ @(v,\ w)] &\bowtie @[\fan t {n + 1}(x,\ v),\ \fan t {n + 1}(y,\ w)]; \\
\fan t m[x,\ y] &\bowtie \fan t n[x,\ y]; \\
\fan t n[x,\ y] &\bowtie \fan u n[x,\ y]; \\
\fan t m[\fan {\langle u + t\rangle} n(x,\ y),\ \fan {\langle u - t\rangle} n(v,\ w)] &\bowtie \fan u n[\fan t {m + 1}(x,\ v),\ \fan t {m + 1}(y,\ w)] \quad (t \neq u,\ m > n).
\end{align*}

The last rule in the list above is central for this paper.
Our intuition for this rule is that the fan with greater number of instances duplicates another fan with different identity.
Here, $\langle u \pm t \rangle$ can be thought of as the left/right clone of $u$ created by $t$.

Translation of $\lambda$-terms to interaction nets is done the same way as in other optimal implementations, except that each bracket, croissant, or delimiter is replaced with just a wire, and each fan is represented as a $\fan x 1$ agent with a unique name $x$.

In order to implement interaction rules in $O(1)$ by time and space, trees can be replaced with hashes, if we represent names as strings.
Specifically, we suggest the following mapping $S(t)$ from trees to hashes, where $\circ$ is string concatenation and $H$ is SHA-256 or any other hash function that is resistant to collisions:
\begin{align*}
S(x) &= H(\text{`$x$'}); \\
S(\langle t_1 + t_2\rangle) &= H(S(t_1) \circ \text{`$+$'} \circ S(t_2)); \\
S(\langle t_1 - t_2\rangle) &= H(S(t_1) \circ \text{`$-$'} \circ S(t_2)).
\end{align*}

Then, both $\langle t_1 \pm t_2\rangle$ operations and equality test for trees become $O(1)$ in time and space.

\section{Prototype}

The following table shows some evaluations of Church numerals and includes benchmarks for the larger examples from \cite{mackie}, providing comparison between CVR and a prototype that implements the solution suggested in this paper.
As usual for such kind of benchmarks, results of the form $N(N_\beta)$ should be read as total of $N$ interactions, of which $N_\beta$ were $\beta$-reductions in the sense of $@ \bowtie \lambda$ interactions.

\begin{center}
\begin{tabular}{|l|l|l|} 
\hline
Term & CVR & Prototype \\ 
\hline
$2\ 2\ 2\ 10\ I\ I$ & $1058(179)$ & $707(67)$ \\
$3\ 2\ 2\ 2\ I\ I$ & $3992(542)$ & $1158(40)$ \\
$10\ 2\ 2\ I\ I$ & $15526(2082)$ & $4282(56)$ \\
$4\ 2\ 2\ 2\ I\ I$ & $983330(131121)$ & $262377(61)$ \\
$2\ 2\ 2\ 2\ 10\ I\ I$ & $4129050(655410)$ & $2359780(198)$ \\
\hline
\end{tabular}
\end{center}

The prototype is being developed in the context of our ongoing project currently focused on obtaining an efficient token-passing optimal reduction with embedded read-back~\cite{termgraph}.
This allowed us to use the prototype on essentially $\lambda K$-terms and compare them against token-passing version of Lambdascope with embedded read-back.
As our main test case, we used a complex $\lambda K$-term representing arithmetical expression $3^3 - (2 + 2)!$ with Church numerals and factorial defined via Turing's fixed point combinator.
With the same waiting construct and read-back mechanisms in use, the benchmarks are as follows: $16024(778)$ for the prototype and $813753(778)$ for Lambdascope.

\section{Further work}

To the best of our knowledge, the suggested mechanism to track identities of fans is a new approach.
This mechanism partially eliminates the need in the oracle.
However, the naive level-tracking part (meant to decide which one of two interacting fans with different identities is active) suffers from counterexamples.
Perhaps, the simplest one is $\omega\ (\lambda x.\omega\ (\lambda y.y\ x))$, where $\omega = \lambda x.x\ x$.
Evaluating that term results in a fan reaching the interface of the net.

Still, we find the suggested approach promising, and believe that it could be possible to improve the level-tracking mechanism in order to block counterexamples and support all $\lambda K$-terms, thus achieving the universal version of Lamping's abstract algorithm.

\printbibliography
\end{document}